# U-NEED: A Fine-grained Dataset for User Needs-Centric E-commerce Conversational Recommendation

Yuanxing Liu
leeeeoliu@gmail.com
Independent
Harbin, Heilongjiang, China

Weinan Zhang*
weinanzhang@gmail.com
Independent
Harbin, Heilongjiang, China

Baohua Dong
baohua.dbh@alibaba-inc.com
Alibaba Group
Hangzhou, Zhejiang, China

Yan Fan
fanyan.fy@alibaba-inc.com
Alibaba Group
Hangzhou, Zhejiang, China

Hang Wang
hwang.rvec@gmail.com
Independent
Harbin, Heilongjiang, China

Fan Feng
fengfan.fengfan@alibaba-inc.com
Alibaba Group
Hangzhou, Zhejiang, China

Yifan Chen
chenyifan.stu@gmail.com
Independent
Harbin, Heilongjiang, China

Ziyu Zhuang
royzhuang1124@gmail.com
Independent
Harbin, Heilongjiang, China

Hengbin Cui
alexcui.chb@alibaba-inc.com
Alibaba Group
Hangzhou, Zhejiang, China

Yongbin Li
shuide.lyb@alibaba-inc.com
Alibaba Group
Hangzhou, Zhejiang, China

Wanxiang Che
wanxiang@gmail.com
Independent
Harbin, Heilongjiang, China

## ABSTRACT
Conversational recommender systems (CRSs) aim to understand the information needs and preferences expressed in a dialogue to recommend suitable items to the user. Most of the existing conversational recommendation datasets are synthesized or simulated with crowdsourcing, which has a large gap with real-world scenarios. To bridge the gap, previous work contributes a dataset E-ConvRec, based on pre-sales dialogues between users and customer service staff in E-commerce scenarios. However, E-ConvRec only supplies coarse-grained annotations and general tasks for making recommendations in pre-sales dialogues. Different from it, we use real user needs as a clue to explore the E-commerce conversational recommendation in complex pre-sales dialogues, namely user needs-centric E-commerce conversational recommendation (UNECR).

In this paper, we construct a *user needs-centric E-commerce conversational recommendation dataset* (U-NEED) from real-world E-commerce scenarios. U-NEED consists of 3 types of resources: (i) 7,698 fine-grained annotated pre-sales dialogues in 5 top categories (ii) 333,879 user behaviors and (iii) 332,148 product knowledge tuples. To facilitate the research of UNECR, we propose 5 critical tasks: (i) pre-sales dialogue understanding (ii) user needs elicitation (iii) user needs-based recommendation (iv) pre-sales dialogue generation and (v) pre-sales dialogue evaluation. We establish baseline methods and evaluation metrics for each task. We report experimental results of 5 tasks on U-NEED. We also report results on 3 typical categories. Experimental results indicate that the challenges of UNECR in various categories are different.

## 1 INTRODUCTION
Conversational recommender systems (CRSs) aim to capture users' information needs and preferences expressed in a dialogue and then recommend appropriate items to users [12]. To facilitate the research of CRSs, most previous work builds datasets via (i) synthesize dialogues based on existing user behaviors [5, 39] (ii) simulate dialogues using crowdsourcing under pre-defined interaction patterns [10, 14, 16, 21, 22]. Based on synthesized and simulated dialogues, previous work archives significant progress in developing CRSs [8, 11, 16, 29, 33, 39, 41]. Nevertheless, the research

---

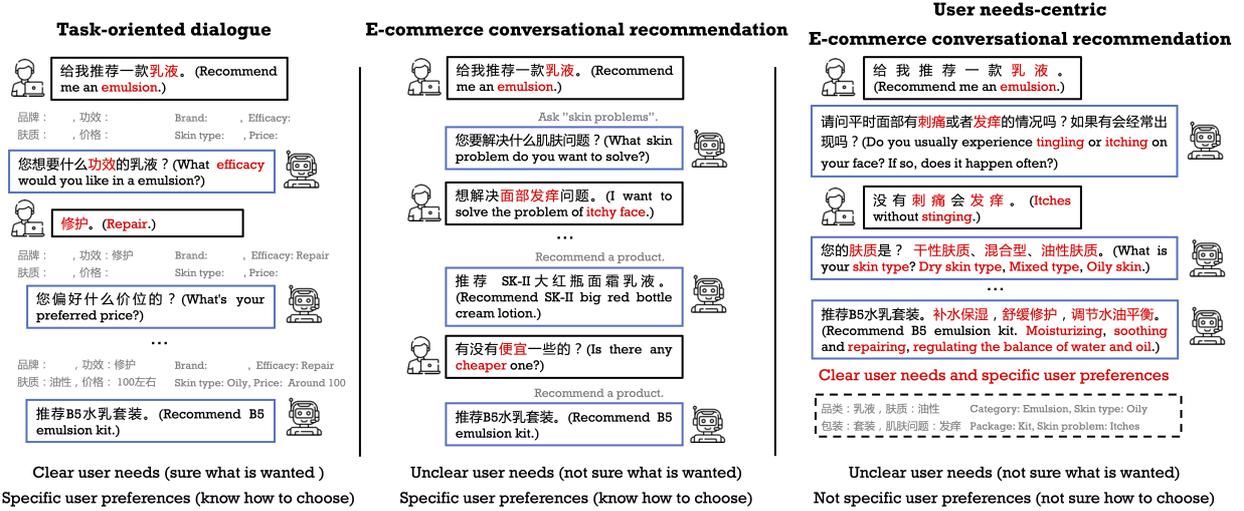

Figure 1: A comparison of task-oriented dialogue (TOD), E-commerce conversational recommendation (ECR) and user needs-centric E-commerce conversational recommendation (UNECR). Different information is highlighted in red. User needs refer to certain features/functionality of a desirable product that a user wants. And user preference refers to the attitude of the user towards an attribute. TOD is dominated by slot values about user needs. ECR mainly focuses on obtaining user preferences and making recommendations. UNECR aims to help the user clarify his or her needs, i.e. from fuzzy to clear. We indicate the main focus of TOD/ECR/UNECR in gray.

on CRSs is limited due to a large gap between natural dialogues and constructed dialogues. To tackle this, Jia et al. [13] propose E-ConvRec, which is built based on natural dialogues between users and customer service staff in E-commerce scenarios. E-ConvRec contains natural dialogues on pre-sales topics, which provides clues to develop CRSs for complex E-commerce scenarios. However, E-ConvRec only provides coarse-grained annotations and tasks for making recommendations in pre-sales dialogues, i.e., user preference recognition, recommendation timing prediction, and personalized recommendation.

**User needs.** In a pre-sales dialogue, a user describes features of the product he/she wants to buy, which we term user needs. By using real user needs rather than simulated user needs as a clue, we explore the E-commerce conversational recommendation in complex pre-sales dialogues, namely user needs-centric E-commerce conversational recommendation (UNECR), in a comprehensive manner. It consists of 5 research topics: (i) understand user needs expressed in utterances (ii) identify attributes that elicit user needs (iii) recommend products that meet user needs (iv) generate responses including attributes that elicit user needs or products that meet user needs and (v) evaluate whether a pre-sale dialogue helps the user clarify his/her needs.

**Resources.** We collect a *user needs-centric E-commerce conversational recommendation dataset* (U-NEED). U-NEED consists of fine-grained annotated pre-sales dialogues, user behaviors and product knowledge tuples. Annotated pre-sales dialogues include 7,698 dialogues and 53,712 utterances derived from a leading Chinese E-commerce platform. For each utterance of pre-sales dialogue, we hire a professional crowdsourcing platform to annotate the action of the speaker, the attributes involved, and the recommended products. For each utterance of annotated pre-sales dialogues, we pay an annotation fee of RMB 0.68. Then, we collect user behaviors and product knowledge tuples. The user behaviors consist of their behaviors before and after pre-sales dialogues, e.g., click and purchase. The product knowledge tuples provide attributes and values. A detailed analysis of U-NEED provides insights for exploring user needs in pre-sales dialogues.

**Tasks.** We propose 5 key tasks about pre-sales dialogues and user needs: (i) Pre-sales dialogue understanding task aims to understand utterances of users and systems to convert nonstandard descriptions into explicit attributes and values. (ii) User needs elicitation task aims to predict attributes that elicit more information about user needs. (iii) User needs-based recommendation task aims at providing recommendations that satisfy existing user needs. (iv) Pre-sales dialogue generation task is designed to generate elicitation questions or recommendation reasons. (v) Pre-sales dialogue evaluation task aims to measure whether the pre-sales dialogue is helpful to the user.

**Benchmarks.** We establish a total of 15 typical baseline methods for the 5 tasks. We report benchmark results of the 5 tasks on U-NEED. Besides, we also report the results of the 5 tasks in 3 top categories. Based on the experimental results, we see that: (i) For the pre-sales dialogue understanding task, all baseline methods have lower performance in the *Phones* category than in other categories. (ii) For both elicitation and recommendation tasks, all baseline methods have lower performance in the *Shoes* category than in other categories. Based on this, we find that the challenges of UNECR in various categories are different.

Table 1: Comparison between U-NEED and existing related datasets. E-commerce(n) means the dataset involves $n$ categories. Natural(c) and Natural(f) refer to the coarse-grained and fine-grained annotations that dialogues have about user needs, respectively. For external resources, UP and KB refer to user profile and knowledge base. UB refers to user behaviors before and after dialogues, while UB(h) refers to historical user behaviors, i.e. before dialogues. Und., Eli., Rec., Gen. and Eval. refer to the 5 aspects of focused tasks, namely, understanding, elicitation, recommendation, generation and evaluation.

| Datasets | Statistic information | | | | | Focused tasks | | | | |
|---|---|---|---|---|---|---|---|---|---|---|
| | #Dialogues | #Utterances | Domain | Dialogue source | External resources | Und. | Eli. | Rec. | Gen. | Eval. |
| FacebookRec [5] | 1M | 6M | Movie | Synthetic | - | | | ✓ | | |
| ReDial [16] | 10,006 | 182,150 | Movie | Simulated | | ✓ | | ✓ | ✓ | |
| CCPE-M [28] | 502 | 11,972 | Movie | Simulated | - | | ✓ | | | |
| GoRecDial [14] | 9,125 | 170,904 | Movie | Simulated | UB(h) | | | ✓ | ✓ | |
| INSPIRED [10] | 1,001 | 35,811 | Movie | Simulated | - | | | ✓ | ✓ | |
| TG-ReDial [39] | 10,000 | 129,392 | Movie | Synthetic | UP | ✓ | | ✓ | ✓ | |
| MGConvRec [35] | 7,615 | 73,971 | Restaurant | Simulated | UB(h) | ✓ | ✓ | ✓ | | |
| OpenDialKG [24] | 15,673 | 91,209 | Movie, book | Simulated | KB | | | ✓ | | |
| DuRecDial [22] | 10,190 | 155,447 | Movie, restaurant | Simulated | UP, KB | | ✓ | ✓ | ✓ | |
| DuRecDial 2.0 [21] | 10,190 | 255,346 | Movie, restaurant | Simulated | UP, KB | | ✓ | ✓ | ✓ | |
| HOOPS [7] | - | 11.6M | E-commerce(4) | Synthetic | KB | | ✓ | ✓ | | |
| E-ConvRec [13] | 25440 | 775,338 | E-commerce | Natural(c) | UP, KB | ✓ | | ✓ | | |
| U-NEED (Ours) | 7,698 | 53,712 | **E-commerce(5)** | **Natural(f)** | **UB**, KB | ✓ | ✓ | ✓ | ✓ | ✓ |

**Contributions.** Main contributions of this paper are as follows:
- We contribute U-NEED, a dataset from E-commerce scenarios. U-NEED consists of 7,698 fine-grained dialogues, 333,879 user behaviors, and 332,148 product knowledge tuples.
- To facilitate research of user needs in pre-sales dialogues, we design 5 tasks (3 for pre-sales dialogue and 2 for user needs). Among them, to our knowledge, we are the first to design an evaluation task when constructing resources for CRSs.
- We establish baselines and report benchmark results for 5 tasks. We release the code and scripts used for the experiments.

## 2 RELATED WORK

The related work lies in two aspects: datasets and tasks. We present a detailed comparison of U-NEED with existing datasets in Table 1.

### 2.1 Datasets for conversational recommendation

To facilitate the research of CRSs, a large amount of resource work has been proposed from three lines.

**Datasets based on synthesized dialogues.** A simple idea is to synthesize dialogues using user behaviors along with pre-defined interaction strategies and utterance templates.[1] Dodge et al. [5] use ratings of MovieLens and natural language templates to synthesize one turn of recommendation. It is then combined with a pair of question and answer to form a 3-turn dialogue, i.e. FacebookRec. Zhou et al. [39] propose TG-ReDial using reviews of Douban Movie.[2] Dialogues of TG-ReDial are synthesized by creating an evolving topic thread that leads from the previous topic to the target topic of recommended movies. In HOOPS, Fu et al. [7] construct a knowledge graph based on Amazon reviews [25], then extract key entities to build user-item interactions, and finally synthesize dialogues based on templates. By synthesizing dialogues, a large amount of data can be constructed to train policy modules and recommendation modules of CRSs. As utterances are structured based on predefined and limited templates, it is hard to train CRSs to generate diverse and persuasive responses using synthetic data.

**Datasets based on simulated dialogues.** Most related work predefines conversational recommendation scenarios and then uses crowdsourcing to simulate dialogues [10, 14, 16, 21, 22, 24, 35]. Among them, Li et al. [16] construct dialogues around movie recommendations, i.e. ReDial, where one worker plays the user and another worker plays the recommender. Dialogues end with a condition that at least 4 movies are mentioned. Then some work supplement ReDial with external resources, e.g. Wikipedia [1], ConceptNet [38] and reviews [23]. Hayati et al. [10] build INSPIRED following an annotation scheme related to recommendation strategies based on social science theories. Liu et al. [22] simulate multi-type dialogues by asking the recommender to proactively lead the dialogue and then make recommendations with consideration of the seeker's interests, instead of the seeker asking for a recommendation from the recommender. With simulated dialogues, we can train CRSs to learn recommendation reasons provided by annotators. To facilitate data collection, the interaction patterns of simulated scenarios tend to be simple and well-defined, which makes the CRS trained on simulated datasets limited to specific scenarios. In addition, the diversity and quality of simulated dialogues are to some extent limited by the habit and knowledge background of annotators.

**Datasets based on natural dialogues.** Synthetic as well as simulated dialogues have a large gap with that in real-world scenarios. Recently, Jia et al. [13] propose a dataset E-ConvRec, which contains dialogues on pre-sales topics between users and customer service staff in E-commerce scenarios. Same as E-ConvRec, we construct U-NEED based on natural dialogues in E-commerce scenarios.

---
[1] MovieLens ratings: https://grouplens.org/datasets/movielens/.
[2] Douban Movie https://movie.douban.com/.

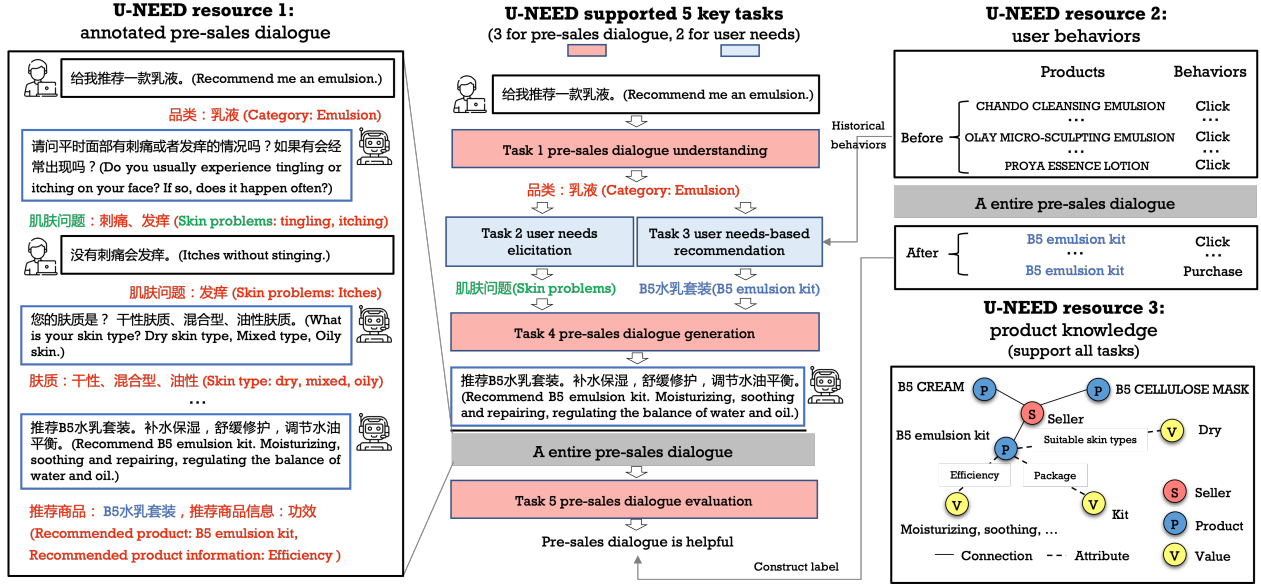

Figure 2: A dialogue example of U-NEED. U-NEED consists of 3 types of resources: annotated pre-sales dialogues, user behaviors, and product knowledge. In red, green, and blue highlight annotations related to user needs. U-NEED supports 5 key tasks in 2 aspects: pre-sales dialogue and user needs. 3 tasks for pre-sales dialogue are represented by red squares. 2 tasks for user needs are represented by blue squares. The middle part also shows connections between the 5 tasks. Gray lines indicate connections between resources and tasks.

Compared with existing datasets, U-NEED has the following strengths: (i) U-NEED provides fine-grained annotations about user needs in pre-sales dialogues. With U-NEED, we can develop E-commerce CRSs for pre-sales scenarios. (ii) U-NEED covers 5 popular categories, providing clues to analyze the needs of users in different categories. (iii) U-NEED provides user behaviors before and after the dialogue, which allows us to explore methods to evaluate CRSs.

## 2.2 Tasks in conversational recommendation

Referring to the general framework of CRSs [8], we summarize the tasks that existing datasets focus on in 5 aspects: understanding, elicitation, recommendation, generation and evaluation.

**Understanding.** We consider the understanding task as capturing the user's needs or preferences from the dialogue context. In ReDial, Li et al. [16] propose a task to infer the sentiment and opinions based on utterances. Xu et al. [35] design a task to predict a user's preference expressed in a conversation as pairs of opinion targets (an item or a value) and their associated sentiment polarities, i.e. positive, negative and neutral. Jia et al. [13] propose the user preference recognition task to identify descriptive words, category words, comparative words and negative words in a user's utterance. Different from previous work, our proposed understanding task not only identifies words related to needs or preferences but also recognizes the product attributes they involve.

**Elicitation.** We consider the elicitation task as identifying information that elicits the user's needs or preferences. Note that strategies about when to ask questions or recommend products [13, 14] are not in the scope of elicitation. In TG-ReDial [39] and DuRecDial [22], this task aims to predict the next topic or goal to lead the dialogue to approach the recommendation target. In HOOPS [7] and MGConvRec [35], the elicitation task is mainly about selecting a question that requests preference over a slot or value. Different from previous work, our proposed elicitation task selects multiple attributes that elicit user needs.

**Recommendation.** We consider the recommendation task as recommending items that satisfy the user's needs or preferences. In most previous work the recommendation task aims to recommend items that users are likely to accept based on the dialogue context [5, 7, 14, 16, 21, 22, 24, 35, 39]. Among them, Jia et al. [13] design personalized recommendation as a ranking task to judge whether the user will buy a candidate product based on user profile, product knowledge base (KB) and dialogue context. Our recommendation task is defined as recommending a product that meets the obtained user's needs.

**Generation.** We consider the generation task as generating a response containing the given information. Most previous work aims to generate dialogue utterances to speak with the seeker in a way like the human speaker [14, 21, 22]. Besides, some previous work aims to incorporate topic [39] or social strategy [10] to the responses to lead the dialogue. Li et al. [16] design the generation task to generate a response that includes the recommended movie.

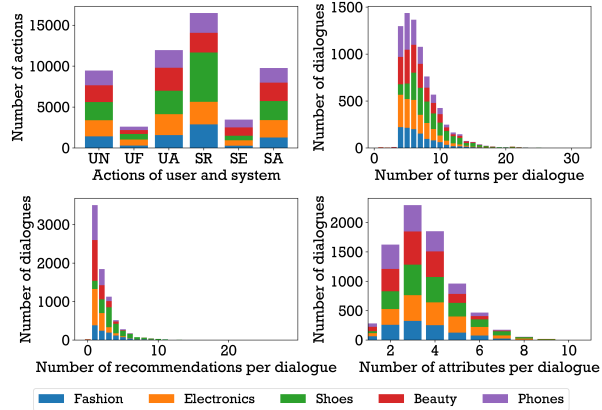

Figure 3: Statistics of annotated pre-sales dialogues. UN, UF and UA are short of *user need*, *user feedback* and *user answer*, respectively. And SR, SE and SA refer to *system recommend*, *system explain* and *system ask*.

Different from previous work, our proposed generation task is designed to generate a response related to attributes identified by the elicitation task or items predicted by the recommendation task.

**Evaluation.** To the best of our knowledge, we are the first to propose an evaluation task when constructing resources for CRSs. We consider the task as evaluating whether the dialogue that the user engaged in with a CRS helps the user to clarify his/her needs.

Compared with existing datasets, to the best of our knowledge, U-NEED is the first E-commerce conversational recommendation dataset that: (i) supports all common tasks in conversational recommendation and (ii) includes an evaluation task that cannot be supported by previous work.

## 3 DATASET

We construct a *user needs-centric E-commerce conversational recommendation dataset* (U-NEED) derived from a leading Chinese E-commerce platform.[3] U-NEED consists of 3 types of resources: (i) annotated pre-sales dialogues (ii) user behaviors and (iii) product knowledge, as shown in Figure 2.

### 3.1 Annotated pre-sales dialogues

*3.1.1 Dialogue collection.* We construct annotated pre-sales dialogues by first collecting utterances from human-human conversations and then performing fine-grained human annotation and desensitization. First, we collect pre-sales dialogues between users and customer service staff from September 15, 2022 to October 15, 2022. To explore pre-sales dialogues in various categories, we collect conversations from 5 categories, i.e. *Beauty*, *Phones*, *Fashion*, *Shoes* and *Electronics*. Inspired by [36], we employ an intent recognition model that is deployed in the production environment to identify actions of speakers and filter dialogues in which the user does not show needs. Then we use a professional crowdsourcing platform appen to perform fine-grained annotation on pre-sales dialogues.[4] We hire 7 employees with undergraduate degrees. 5 of them annotate the action of speakers, the attributes involved and recommended products. The details of annotation tasks can be found here.[5] The remaining 2 employees are responsible for quality control. We give appen 9,545 dialogues and then receive 7,698 qualified annotated pre-sales dialogues. For each qualified utterance of annotated pre-sales dialogues, we pay an annotation fee of RMB 0.68.

[3]https://www.taobao.com/

Table 2: Toy examples of user behavior tuples.

| Product id | User id | Behavior id | Timestamp |
| --- | --- | --- | --- |
| 1 | 2 | 3 | 2022-05-05 16:25:25 |
| 1 | 2 | 5 | 2022-05-05 17:26:55 |

Table 3: Toy examples of product knowledge tuples.

| Product id | Attribute | Value | Seller id |
| --- | --- | --- | --- |
| 1 | 价格区间 (Price range) | 中(Middle) | 2 |
| 1 | 功效 (Effect) | 抗皱 (Anti-wrinkle) | 2 |
| 1 | 品牌 (Brand) | 玉兰油 (OLAY) | 2 |

*3.1.2 Dialogue analysis.* We collect 7,698 fine-grained annotated pre-sales dialogues, which consist of 1662, 1513, 1135, 1748, and 1640 dialogues in *Beauty*, *Phones*, *Fashion*, *Shoes* and *Electronics* categories respectively. As shown in Figure 3, we present statistics of annotated pre-sales dialogues in four critical aspects.

**Actions of user and system.** In the upper left part of Figure 3 we present the distribution of actions of users and systems in annotated pre-sales dialogues across categories. The y-axis denotes the number of occurrences of the corresponding action in all dialogues. We see that: (i) Among the six pre-defined actions, SR appears the most times, followed by UA, SA, and UN. UF and SE appear the least times. (ii) The number of different actions in various categories is roughly equal. However, SR appears more frequently in the *Shoes* and *Fashion* categories than in the remaining 3 categories.

**Number of turns.** In the upper right part of Figure 3 we draw the number of dialogues of various lengths across categories. We observe that: (i) The number of turns in dialogues is mainly in the range of 5 to 10. This indicates that short dialogues may be appropriate in real scenarios. (ii) Besides, the proportion of *Shoes* gradually increases as the number of turns increases. This may be due to the fact that it is difficult to describe needs and preferences related to shoes, so a higher number of turns are required.

**Number of recommendations.** In the lower left part of Figure 3 we statistics the number of recommendations in annotated pre-sales dialogues. We see that: (i) Dialogues where only 1 product is recommended account for over one-third of all dialogues. (ii) However, in the *Shoes* category, the dialogues include more recommendations.

**Number of different attributes.** In the lower right part of Figure 3 we statistics the number of attributes involved in annotated pre-sales dialogues. We see that: (i) Dialogues involving 3 different attributes have the highest number. (ii) The proportion of the number of different attributes involved in the 5 categories is even.

[4]https://www.appen.com.cn/
[5]https://github.com/LeeeeoLiu/U-NEED/blob/main/annotation.md

## 3.2 Related resources

*3.2.1 User behaviors.* For each user in the annotated pre-sales dialogues, we collect his/her behaviors to analyze whether the pre-sales dialogue helps the user. We use a tuple $(p, u, b, t)$ to denote that user $u$ has a behavior $b$ on product $p$ at time $t$. Table 2 shows 2 toy examples of user behaviors tuples. Each row of the behavior data contains the user id, product id, behavior id and timestamp of the behavior. Behavior is one of click, buy, favorite and add to cart. In practice, we collect the user behavior on the day when the user has a pre-sales dialogue with the customer service staff. We then perform a strict desensitization operation on the collected behaviors to make them untraceable.

We collect 333,879 tuples of user behaviors, which involve 7,620 users and 68,027 products. Based on the timestamp of the collected dialogue, we can divide the user's behaviors into two parts. The user behaviors before the dialogue starts, which we call user historical behaviors. The user behaviors after the dialogue starts are regarded as user follow-up behaviors. With user historical and follow-up behaviors, we can construct a label to evaluate CRSs.

*3.2.2 Product knowledge.* We collect information about products involved in annotated pre-sales dialogues. We use a tuple $(p, a, v, s)$ to denote that seller $s$ has product $p$ and the value of attribute $a$ of product $p$ is $v$. Table 3 shows 3 toy examples of the product knowledge tuples. Each row consists of product id, attribute, value, and seller id.

We collect 332,148 product knowledge tuples, which involve 68,954 products and 13,122 sellers. With this product knowledge, we can construct a heterogeneous graph [26] to support future research in pre-sales dialogues.

## 3.3 Research topics

We outline 5 promising research topics (RT) of user needs-centric E-commerce conversational recommendation (UNECR) based on the in-depth analysis of U-NEED and the resources it provides.

*RT1: Understand user needs expressed in utterances.* A notable difference between real-world scenarios and simulated scenarios is the user group. U-NEED includes numerous various descriptions of users about their needs. In the *Phones* category, there are more than 370 different ways to express needs regarding functionality.

*RT2: Identify attributes that elicit user needs.* In real-world scenarios, users have vague needs and are unfamiliar with product attributes. Experienced customer service staff can quickly guide users to clarify their needs. U-NEED provides labels of the attributes that the staff asks in each turn, which provide clues to investigate the relationship between user needs and product attributes.

*RT3: Recommend products that meet user needs.* Recommendations are more challenging in real-world scenarios. Most users have limited patience, and customer service staff must recommend appropriate products with little information about needs or preferences. Using U-NEED, we can explore the real-world recommendations provided by the customer service staff.

*RT4: Generate responses that contain attributes that elicit user needs or items that meet user needs.* CRSs need to transform given information into a way that the user can easily understand. For example, when customer service staff asks for preferences about "skin problem", he/she would express it as "tingling or itching?".

*RT5: Evaluate whether a pre-sale dialogue helps to clarify user needs.* U-NEED contains user behaviors before and after the pre-sale dialogue, which allows us to construct labels for evaluating CRSs and explore the impact of pre-sale dialogues.

## 4 TASKS

To facilitate research on the 5 topics, we design 5 key tasks.

## 4.1 Pre-sales dialogue understanding (task 1)

For RT1, we design a task for pre-sales dialogue understanding. The task aims to understand utterances by: (i) identifying the attributes that are related to user needs and (ii) extracting the corresponding preferences.

Given an utterance $x$, e.g. "Are there any hydration and moisturizing kits?", a model $\mathcal{M}_u$ aims to predict semantic frames $U = \{(a_i, v_i)\}$, i.e. { ("Efficacy", "Hydration and moisturization"), ("Package", "Kits") }, included in $x$. Formally, the task is denoted as:

$$U = \mathcal{M}_u(x). \quad (1)$$

## 4.2 User needs elicitation (task 2)

For RT2, we design a task for user needs elicitation, which aims to select attributes that can elicit more information about user needs.

Given pre-sales dialogue context $X = \{x_i\}$ and contextual semantic frames $C = \{U_i\}$ of $X$, a model $\mathcal{M}_a$ aims to select the attributes $A = \{a_i\}_{i=1}$ that elicit information about user needs. Formally, the task of user needs elicitation is denoted as:

$$A = \mathcal{M}_a(X, C). \quad (2)$$

## 4.3 User needs-based recommendation (task 3)

For RT3, we design a task for user needs-based product recommendation, which aims to recommend products that meet explicit user needs expressed in the pre-sales dialogue and implicit user needs, i.e. user behaviors before the pre-sales dialogue.

Given pre-sales dialogue context $X = \{x_i\}$ and user historical behaviors $B = \{(p_i, b_i)\}$ a model $\mathcal{M}_r$ aims to recommend products $P = \{p_i\}$ meets both explicit and implicit user needs. Formally, the task is denoted as:

$$P = \mathcal{M}_r(X, B). \quad (3)$$

## 4.4 Pre-sales dialogue generation (task 4)

For RT4, we design a task for pre-sales dialogue generation, which aims to generate a pre-sales dialogue response containing the given information about user needs.

Given pre-sales dialogue context $X = \{x_i\}$, products $P = \{p_i\}$ that meet user needs and attributes $A = \{a_i\}$ that can elicit the user needs, a model $\mathcal{M}_g$ aims to generate a response $Y = \{w_i\}$. Formally, the task is denoted as:

$$Y = \mathcal{M}_g(X, A, P). \quad (4)$$

$A$ or $P$ may be $\emptyset$, which means there is no suitable attribute/product.

## 4.5 Pre-sales dialogue evaluation (task 5)

For RT5, we design a task for pre-sales dialogue evaluation, which aims to measure whether the pre-sales dialogue is helpful.

Table 4: Performance of baseline methods on the proposed 5 tasks in 3 typical categories: *Beauty*, *Fashion* and *Shoes*. To present results properly, we abbreviate some evaluation metrics: P and R refer to Precision and Recall. H@K and M@K refer to Hit@K and MRR@K. Info. and Rel. refer to informativeness and relevance. $\kappa$ is the average pairwise Cohen's kappa coefficient between annotators. PCC, SCC and Cos. refer to the Pearson correlation coefficient, Spearman correlation coefficient and cosine similarity. The best results are highlighted in bold.

| | Beauty | | | Shoes | | | Phones | | | All 5 categories | | |
|---|---|---|---|---|---|---|---|---|---|---|---|---|
| | Task 1: pre-sales dialogue understanding | | | | | | | | | | | |
| Methods | P | R | F1 | P | R | F1 | P | R | F1 | P | R | F1 |
| Bert [4] | 0.5355 | 0.6284 | 0.5782 | 0.5851 | 0.7020 | 0.6382 | 0.4212 | 0.5384 | 0.4726 | 0.4549 | 0.5652 | 0.5041 |
| Bert+CRF [31] | 0.6731 | 0.6802 | 0.6766 | 0.7302 | 0.7703 | 0.7497 | 0.5620 | 0.5923 | 0.5768 | 0.6688 | 0.6530 | 0.6608 |
| Bert+BiLSTM+CRF [3] | **0.7282** | **0.7481** | **0.7380** | **0.7870** | **0.8101** | **0.7984** | **0.6701** | **0.6990** | **0.6843** | **0.6892** | **0.6875** | **0.6884** |
| | Task 2: user needs elicitation | | | | | | | | | | | |
| | P | R | F1 | P | R | F1 | P | R | F1 | P | R | F1 |
| DiaMultiClass [18] | 0.4037 | **0.7228** | **0.5054** | 0.3361 | **0.4131** | **0.3423** | **0.4534** | **0.5212** | **0.4585** | 0.3222 | **0.4966** | **0.3662** |
| DiaSeq [18] | **0.4761** | 0.4272 | 0.4424 | **0.3992** | 0.3305 | 0.3498 | 0.4414 | 0.3789 | 0.3966 | **0.3555** | 0.2996 | 0.3153 |
| | Task 3: user needs-based recommendation | | | | | | | | | | | |
| | H@10 | H@50 | M@50 | H@10 | H@50 | M@50 | H@10 | H@50 | M@50 | H@10 | H@50 | M@50 |
| Bert [4] | **0.2985** | **0.3938** | **0.1532** | **0.1014** | **0.2550** | **0.0384** | 0.4275 | 0.7174 | 0.2086 | **0.1593** | **0.3310** | **0.0660** |
| SASRec [15] | 0.1477 | 0.3046 | 0.0378 | 0.0722 | 0.1720 | 0.0344 | 0.4022 | **0.7246** | 0.1695 | 0.1400 | 0.2747 | 0.0585 |
| TG-CRS [39] | 0.1754 | 0.2585 | 0.0778 | 0.0568 | 0.1521 | 0.0231 | **0.4565** | 0.6920 | **0.2145** | 0.1470 | 0.2500 | 0.0655 |
| | Task 4: pre-sales dialogue generation | | | | | | | | | | | |
| | Dist-4 | Info. | Rel. | Dist-4 | Info. | Rel. | Dist-4 | Info. | Rel. | Dist-4 | Info. | Rel. |
| Transformer [34] | 0.3714 | **1.2767** | 0.5267 | 0.3718 | 1.0400 | 0.9600 | **0.4037** | **1.2267** | 0.9467 | 0.2806 | **1.1567** | 0.8800 |
| GPT-2 [27] | **0.3811** | 0.6433 | 0.2767 | 0.3357 | 0.5000 | 0.3767 | 0.3266 | 0.7633 | 0.4700 | **0.2905** | 0.5700 | 0.4267 |
| KBRD [1] | 0.2808 | 1.2233 | 0.6133 | 0.4051 | **1.0933** | 1.0467 | 0.4157 | 1.0900 | 0.9867 | 0.2233 | 1.1367 | 0.9167 |
| NTRD [19] | 0.3275 | 1.1500 | **0.6867** | **0.4635** | 1.0100 | **1.0933** | 0.3775 | 1.0400 | **1.0567** | 0.2489 | 1.0033 | **0.9900** |
| $\kappa$ | - | 0.7162 | 0.3947 | - | 0.8431 | 0.6503 | - | 0.6862 | 0.5759 | - | 0.7523 | 0.4958 |
| | Task 5: pre-sales dialogue evaluation | | | | | | | | | | | |
| | PCC | SCC | Cos. | PCC | SCC | Cos. | PCC | SCC | Cos. | PCC | SCC | Cos. |
| DEB [30] | **0.1642** | **0.1628** | **0.9327** | **0.1504** | **0.1963** | 0.9097 | **0.2678** | **0.2815** | **0.9366** | **0.1617** | **0.1864** | 0.9212 |
| P-value | <0.0299 | <0.0313 | - | <0.0416 | <0.0076 | - | <0.0015 | <0.0008 | - | <6e-06 | <1e-07 | - |
| Bert-RUBER [9] | 0.0901 | 0.1133 | 0.9218 | 0.0916 | 0.1157 | **0.9219** | 0.0900 | 0.1141 | 0.9218 | 0.0742 | 0.1092 | **0.9214** |
| P-value | <0.0126 | <0.0017 | - | <0.0111 | <0.0013 | - | <0.0126 | <0.0015 | - | <0.0398 | <0.0024 | - |

Given an entire pre-sales dialogue $D = \{x_i\}$, a model $\mathcal{M}_e$ aims to predit a rating $r$ of pre-sales dialogue $D$. Formally, the task of pre-sales dialogue evaluation is denoted as:

$$r = \mathcal{M}_e(D). \tag{5}$$

## 5 EXPERIMENTS AND RESULTS

Table 4 presents the benchmark results for 5 tasks. We introduce data preparation, baseline methods, evaluation metrics, and results and analysis of each task separately in the following subsections.

Following [20], we use the timestamps of annotated pre-sales dialogues to split the entire dataset into training set, validation set, and test set. U-NEED spans 30 days, and we utilize dialogues of the first 24 days as the training set, dialogues of the last 3 days as the test set, and the rest as the validation set. We then construct the inputs and labels required for each task. To facilitate the reproducibility of the experimental results, we release the code and scripts used in data pre-processing and experiments.[6]

### 5.1 Benchmark for task 1

*5.1.1 Data preparation.* Based on the manually annotated attributes and preferences, we construct a label for each word in an utterance using BIO tagging.[7] Given an utterance "Are there any hydration and moisturizing kits?" and its annotated attributes and preferences { ("Efficacy", "Hydration and moisturization"), ("Package", "Kits") }, the label is constructed as "Are/O there/O any/O hydration/B-Beauty-Efficacy and/I-Beauty-Efficacy moisturizing/I-Beauty-Efficacy kits/B-Beauty-Package ?/O". We build 41,592 training samples, 6,251 validation samples and 2,874 test samples.

*5.1.2 Baseline methods.* We select 3 models that are effective and used in the production environments of E-commerce scenarios.

---
[6]https://github.com/LeeeeoLiu/U-NEED
[7]https://en.wikipedia.org/wiki/Inside-outside-beginning_(tagging)

**Bert [4]** predicts the probability of each tag based on the representation of the input utterance.

**Bert+CRF [31]** predicts the probability of each tag considering the representation of the input utterance as well as the sequential relationship between predicted tags.

**Bert+BiLSTM+CRF [3]** considers bi-directional information after the encoding of bert to obtain a better representation of the input utterance. It also considers the sequential relationship between predicted tags.

*5.1.3 Evaluation metrics.* Following [3], we adopt precision, recall, and F1 score as evaluation metrics.

**Precision** is the ratio of the number of correctly identified tags to the number of all identified tags.

**Recall** is the ratio of the number of correctly identified tags to the number of originally correct tags.

**F1 score** is calculated as the summed average of the precision and recall.

*5.1.4 Results and analysis.* From Table 4, we observe that in all 5 categories, the Bert+BiLSTM+CRF model achieves the best results. The Bert+BiLSTM+CRF model achieves an F1 value of 0.7984 in the *Shoes* category, while in the *Phones* category the F1 value is only 0.6843. By analyzing utterances, we find that the expression of users' needs for shoes is relatively similar across dialogues, while there are large differences in the *Phones* category. Modeling relations between users' irregular expressions and fine-grained product attributes may be a worthwhile research topic for future work.

## 5.2 Benchmark for task 2

*5.2.1 Data preparation.* We construct samples based on utterances of the customer service staff asking about preferences on attributes. Take the annotated pre-sales dialogue in Figure 2 as an example. In the second turn of the dialogue, the system is asking the user a question about "Skin problem". The input is the dialogue context and the recognized user needs, i.e. "Category: Emulsion". The label is the attribute that the customer service is asking about in this turn, i.e. "Skin problem". Note that the label may contain more than one attribute. For example, "Do you have any requirements for the function of the washing machine? What is the budget?" includes both "Function" and "Price". We build 7,432 training samples, 1,108 validation samples and 1,042 test samples.

*5.2.2 Baseline methods.* Inspired by [18], we set up 2 baseline models from multi-label classification and sequence generation.

**DiaMultiClass [18]** encodes the input text and then predicts the probability of each attribute. In the inference stage, we set a threshold of 0.5 to output the prediction results.

**DiaSeq [18]** utilizes a GRU-based decoder to predict attribute combinations.

*5.2.3 Evaluation metrics.*

**Precision** is the ratio of the number of correctly selected attributes to the number of all selected attributes.

**Recall** is the ratio of the number of correctly selected attributes to the number of originally correct attributes.

**F1** is obtained by calculating the summed average of the precision and recall.

*5.2.4 Results and analysis.* From Table 4, we see that DiaMultiClass achieves better performance than DiaSeq in Recall in all categories. Besides, we observe a phenomenon in *Beauty* and *Phones* categories, i.e. DiaMultiClass does not perform as well as DiaSeq in terms of Precision but outperforms DiaSeq in both Recall and F1 score. In terms of *Shoes*, we see two methods that achieve comparable results in terms of F1 score.

## 5.3 Benchmark for task 3

*5.3.1 Data preparation.* We construct samples based on utterances of the customer service staff recommending products in the pre-sales dialogues. Take the annotated pre-sales dialogue in Figure 2 as an example. In the final turn of the dialogue, the system is recommending a product "B5 emulsion kit". The input is the dialogue context and the user's historical behaviors. The label is the recommended product in this turn, i.e. "B5 emulsion kit". Note that customer service staff may recommend multiple products at one time. Following [37, 38, 40], we process multiple recommendations as multiple samples with the same input. We build 13,671 training samples, 2,064 validation samples and 1,864 test samples.

*5.3.2 Baseline methods.* We consider 3 widely applied methods that use different types of information to make recommendations.

**Bert [4]** encodes the dialogue context and then makes recommendations.

**SASRec [15]** employs a self-attention mechanism to capture long-term semantics in users' historical behaviors to generate recommendations.

**TG-CRS [39]** fuses the representation of dialogue context and users' historical behaviors to make recommendations.

Here we modify TG-CRS to fit our task by using the bert model to encode dialogue context and employing users' historical behaviors as topic threads.

*5.3.3 Evaluation metrics.* We adopt 2 widely used metrics for evaluating top-$K$ ($K = 1, 10, 50$) recommendation performance [17, 40].

**Hit@$K$** is the fraction of relevant items that are returned in the top-$K$ ranking out of all relevant items.

**MRR@$K$** is computed as the average of the reciprocal rank of the items that are returned in the top-$K$ ranking. If an item is not returned in top-$K$ rank, its reciprocal rank is 0.

*5.3.4 Results and analysis.* From Table 4, we observe a large gap in recommendations across categories. In the *Phones* category, the performance achieved by each model is significantly higher than those in other categories. Specifically, TG-CRS performs well in the *Phones* category, where it achieves the best results on both H@10 and M@50. In the other three categories, bert achieves the best results on all metrics.

## 5.4 Benchmark for task 4

*5.4.1 Data preparation.* We construct samples using responses from the customer service staff in the pre-sales dialogues. The input is the pre-sales dialogue context and the ground truth is the response from customer service response. We build 15,339 training samples, 1,958 validation samples and 1,932 test samples.

*5.4.2 Baseline methods.* We set up 4 baseline methods. 2 of them are widely used in response generation, and 2 are methods of generating responses that contain recommended items in CRSs.

**Transformer [34]** applies a transformer-based encoder-decoder framework to generate proper responses.
**GPT-2 [27]** is a pre-training text generation model and fine-tuned on U-NEED dataset.
**KBRD [1]** applies a transformer with enhanced modeling of word weight based on knowledge graphs.
**NTRD [19]** generates response templates with special tokens and then replaces the special token with the recommended item.

*5.4.3 Evaluation metrics.* We adopt 3 commonly used metrics for evaluating generation performance [40].

**Distinct@4** is computed as the average of the fraction of the number of distinct 4-grams out of the number of all 4-grams in a response, which measures the diversity of generated responses.
**Informativeness** is computed as the average of the informativeness degree of all generated responses. The score is in $\{0, 1, 2\}$, and we compute an average of scores from all annotators as the final score.
**Relevance** is computed as the average of the relevance degree of all generated responses. The annotators rate to what degree a generated response contains product information related to that in the ground truth.

Informativeness and Relevance are for human evaluation, and we recruit 12 annotators to evaluate 100 randomly selected generated responses in *Beauty*, *Shoes*, *Phones* and all 5 categories. The score is in $\{0, 1, 2\}$, and we compute an average of scores from all annotators as the final score. We calculate the Fleiss's kappa [6] to measure the inter-annotator agreements.

*5.4.4 Results and analysis.* From Table 4, we see that KBRD and NTRD outperform Transformer and GPT-2 in terms of relevance across all categories. This is because the former explicitly incorporates product information in the response generation process. Besides, we observe that the relevance scores are much lower in the *Beauty* category than in other categories, and the annotators appear to have different judgments. This is because most of the pre-sales conversations in the *Beauty* category are about the efficacy of the product. This usually involves a lot of non-standard descriptions of needs, making pre-sales dialogue generation in the *Beauty* category challenging.

## 5.5 Benchmark for task 5

*5.5.1 Data preparation.* Inspired by [2], we construct objective ratings based on correlations between user follow-up behaviors and products that are recommended in the pre-sales dialogue. We build 5,517 training samples, 821 validation samples and 767 test samples.

*5.5.2 Baseline methods.* We select 2 commonly used baseline methods in dialogue evaluation.

**DEB [30]** is pre-trained on 727M Reddit dialogues and then fine-tuned on the DailyDialog++ dataset.
**Bert-RUBER [9]** performs 3 changes to a widely applied evaluation model RUBER [32]: word2vec embeddings are substituted with bert embeddings, Bi-RNNs are replaced with pooling strategies, and the ranking loss is replaced with a classification loss.

We use pre-sales dialogues to fine-tune bert models for DEB and Bert-RUBER.

*5.5.3 Evaluation metrics.* Following [9], we adopt 3 commonly used metrics for evaluating the performance of baseline models.
**Pearson** refers to the Pearson correlation coefficient, which measures a linear correlation between two ordinal variables.
**Spearman** refers to the Spearman correlation coefficient, which measures any monotonic relationship.
**Cosine similarity** computes how much the scores produced by models are similar to scores constructed based on user behaviors.

*5.5.4 Results and analysis.* From Table 4, we see that DEB outperforms bert-RUBER in all correlation coefficient metrics. DEB performs better in the *Phones* category than it does in other categories. It may be because the criteria for the superiority and inferiority of products in the *Phones* category are clearer, such as large screens and specific photographic pixel values. While the criteria for *Shoes* and *Beauty* are more complex, involving product appearance and individual needs.

## 6 DATASET ACCESS

We adhere to strict policies and rules to protect the privacy of users in collecting and processing pre-sales dialogues. Since U-NEED is collected from real-world scenarios, accessing U-NEED requires filling out a form.[8] We will send you U-NEED by e-mail when your application is approved. Note that U-NEED is only for research purposes. Without permission, it may not be used for any commercial purposes or distributed to others. We provide 5 publicly available toy examples.[9] These examples are appropriately modified to show the features of pre-sales dialogues but do not contain factual information.

## 7 CONCLUSIONS

In this paper, we focused on the leading role of real user needs in pre-sales dialogues, which is termed as UNECR. To facilitate research of UNECR, we proposed a fine-grained dataset U-NEED, which consists of annotated pre-sales dialogues, user behaviors and product knowledge tuples, covering 5 top categories. Based on U-NEED, we proposed 5 key tasks and established 15 baseline methods for pre-sales dialogue and user needs. Based on experimental results and analysis of 5 tasks on U-NEED, we found that: (i) There is a big difference in the way users describe their needs in different categories. In the *Fashion* category descriptions of user needs are more diverse than that in the *Shoes* category. (ii) Number of recommendations varies in different categories. In the *Shoes* category, the number of recommendations in pre-sales dialogues is greater than in other categories.

U-NEED has huge worth in three aspects: new significance, practical application, and future research. (i) U-NEED shows that user needs are an essential and effective entry point for exploring conversational recommendation tasks in E-commerce scenarios. (ii) With U-NEED we can develop CRSs that can be applied to real-world scenarios. (iii) U-NEED supports future research on pre-sales dialogue and user needs.

---
[8]https://github.com/LeeeeoLiu/U-NEED/blob/main/dataset_access.md
[9]https://github.com/LeeeeoLiu/U-NEED/blob/main/examples.md

## ACKNOWLEDGMENTS

We thank the anonymous reviewers for their helpful comments. This work is supported by the Science and Technology Innovation 2030 Major Project of China (No. 2020AAA0108605) and National Natural Science Foundation of China (No. 62076081, No. 61772153, and No. 61936010).